\documentclass[9pt, conference]{IEEEtran}
\IEEEoverridecommandlockouts

\usepackage{cite}
\usepackage{amsmath,amssymb,amsfonts}
\usepackage{algorithm}
\usepackage{comment}
\usepackage{graphicx}
\usepackage{textcomp}
\usepackage{booktabs}
\usepackage{array}
\usepackage{hyperref}
\usepackage{xcolor}
\usepackage{threeparttable}
\usepackage{multicol, multirow}
\usepackage{bm}
\usepackage{subfigure}

\def\L{{\cal L}}

\usepackage{enumitem}
\usepackage{xcolor}
\usepackage{algpseudocode}
\usepackage{pifont}
\newcommand{\cmark}{\ding{51}}%
\newcommand{\xmark}{\ding{55}}%

\def\D{{\cal D}}

\def\T{{\cal T}}

\definecolor{darkgreen}{RGB}{0,100,0}

\newcommand{\subrule}{\specialrule{0.25pt}{2pt}{2pt}}

\def\BibTeX{{\rm B\kern-.05em{\sc i\kern-.025em b}\kern-.08em
    T\kern-.1667em\lower.7ex\hbox{E}\kern-.125emX}}
\begin{document}

\title{CGaLore: Curvature-Guided GaLore for Memory-Efficient Continual Adaptation of ASR Foundation Models}

\author{\IEEEauthorblockN{Steven Vander Eeckt} \IEEEauthorblockA{\textit{Department of Electrical Engineering ESAT/PSI} \\\textit{KU Leuven}\\ Leuven, Belgium \\ steven.vandereeckt@esat.kuleuven.be} \and \IEEEauthorblockN{Hugo Van hamme} \IEEEauthorblockA{\textit{Department of Electrical Engineering ESAT/PSI} \\ \textit{KU Leuven}\\ Leuven, Belgium \\ hugo.vanhamme@esat.kuleuven.be}}



\maketitle

\begin{abstract}
Automatic speech recognition models suffer from catastrophic forgetting when adapted to new domains, accents, or downstream tasks. This problem becomes increasingly important with the growing use of speech foundation models, where adaptation should be both memory-efficient and safe, preserving the broad capabilities learned during pretraining. Gradient Low-Rank Projection (GaLore) has recently been proposed as a memory-efficient fine-tuning method that keeps model parameters full-rank while reducing the optimizer memory through low-rank gradient projection. However, GaLore does not account for catastrophic forgetting. We propose Curvature-Guided GaLore (CGaLore), which incorporates old-task curvature information when selecting the low-rank projection bases. Specifically, CGaLore filters current-task gradients using Kronecker-factored approximate curvature from previous tasks before computing the gradient subspace. Our experiments show that CGaLore enables effective adaptation while alleviating forgetting, outperforming state-of-the-art continual learning baselines. Extensive ablation studies confirm the practical applicability of CGaLore.\end{abstract}


\section{Introduction}

Automatic speech recognition (ASR) has increasingly moved toward large foundation models: general-purpose models trained on large and diverse speech corpora, often across many languages, domains, and acoustic conditions. Examples include OWSM~\cite{owsm}, Whisper~\cite{whisper}, OWSM v4~\cite{owsmv4}, Granite Speech~\cite{granite-speech-4.1-2b}, Qwen3-ASR~\cite{Qwen3-ASR}, and recent Canary and Parakeet models~\cite{sekoyan2025canary1bv2parakeettdt06bv3efficient}. Because of their broad pretraining, these models provide strong starting points for downstream adaptation, making it possible to obtain excellent performance on specific domains, accents, speakers, or tasks.

Adapting such models, however, raises two important challenges. First, foundation models contain many parameters, making full fine-tuning computationally intensive. This has motivated parameter-efficient fine-tuning methods, such as Low-Rank Adaptation (LoRA)~\cite{lora}. Second, adaptation can lead to catastrophic forgetting~\cite{catastrophicforgetting}: while learning a new task, the model may overwrite previously acquired knowledge. This is undesirable for foundation models, where adaptation should add new expertise without sacrificing the broad capabilities obtained during pretraining. The objective is therefore continual learning (CL): adapting to new tasks while preserving performance on previously learned ones.

The combination of efficiency and CL gives rise to parameter-efficient continual learning (PECL)~\cite{zhao2024sapt}, where the goal is to adapt large pretrained models efficiently and without forgetting. In natural language processing (NLP) and computer vision, PECL has received increasing attention, with several methods building on LoRA and related low-rank adaptation techniques~\cite{zhao2024sapt,bilora,qiao2026merge,luo2026keeplora,ewclora}. In ASR, however, PECL remains comparatively underexplored. \cite{xu24h_interspeech} combine orthogonal LoRA \cite{wang-etal-2023-orthogonal} with AdaLoRA~\cite{adalora}, while \cite{ugan25_interspeech} apply weight averaging~\cite{weight_averaging} to task-specific LoRA modules. These approaches do not directly address forgetting with respect to the initial foundation model; while task-specific adapter methods~\cite{adapters, eeckt_adapters} typically require the task identity to be known at inference time. Finally, recent Singular Value Decomposition (SVD)-based PECL methods for ASR constrain adaptation to selected singular subspaces~\cite{wang2025ssvd,cssvd}.

More recently, GaLore~\cite{galore} has offered an alternative to LoRA-based fine-tuning. Instead of freezing the pretrained weights and learning additional low-rank modules, GaLore keeps the model parameters full-rank but projects full-rank gradients into a much lower-dimensional subspace before applying the optimizer. The optimizer update is then projected back to the original parameter space. Since optimizers often store parameter-shaped states~\cite{adam,adamw,adagrad}, this substantially reduces optimizer-state memory while still allowing full-rank parameter updates over time. Several recent works have extended GaLore. Q-GaLore~\cite{zhang2024qgalore} further reduces memory through quantization and layer-adaptive subspace updates, while GaLore~2~\cite{galore2} accelerates basis updates using randomized SVD~\cite{halko2011finding}. Its memory-efficient properties make GaLore a promising basis for memory-efficient continual learning. Recent work in NLP has explored this direction by using optimizer first-order moments~\cite{wang-etal-2025-continual} or historical gradients~\cite{cheng2026continuous} to reduce interference with old tasks. However, they require information collected during the initial foundation model training process, typically unavailable when adapting an already released ASR foundation model to a downstream task.

In this work, we propose Curvature-Guided GaLore (CGaLore), a memory-efficient CL method for safe adaptation of foundation models. CGaLore uses Kronecker-factored approximate curvature (KFAC) estimated on old-task data with the initial foundation model to approximate the old-task Hessian. This curvature information is used to filter the current-task gradient before computing the projection bases. As a result, the low-rank optimizer subspace is guided toward directions that are relevant for the new task while being less harmful to previous tasks. CGaLore therefore preserves the main advantage of GaLore---memory-efficient full-rank adaptation---while explicitly biasing the adaptation toward directions that reduce forgetting.

Our contributions are as follows. First, we study memory-efficient CL for ASR foundation-model adaptation in the practical setting where a pretrained model is available, but pretraining gradients, optimizer states, and full pretraining data are not. Second, we propose CGaLore, a curvature-guided extension of GaLore that uses old-task KFAC to guide the low-rank optimizer subspaces during training. Third, we show that CGaLore improves the learning--retention trade-off over state-of-the-art baselines, substantially reducing GaLore's forgetting while preserving its efficiency. Finally, we analyze the role of KFAC estimation data, projection structure, and post-hoc inverse-Hessian correction, showing that CGaLore can work with limited old-task data and can also reduce forgetting on held-out old tasks.

\section{Problem Formulation \& Background}
\subsection{ASR Model \& Notation}
\label{subsec:notation}
We consider an ASR (foundation) model with parameters $\bm\theta \in \mathbb{R}^{N}$ that maps a speech utterance $\bm X \in \mathbb{R}^{F \times d_s}$ to a target sequence $\bm y$ of $w$ output tokens. $F$ denotes the number of acoustic frames and $d_s$ the feature dimension per frame. Given a paired training example $(\bm X,\bm y)$, the model is optimized by minimizing the ASR loss $\L(\bm X,\bm y;\bm\theta)$. The adaptation methods considered in this work operate on linear weight matrices of the model. We denote such a matrix by $\bm W \in \mathbb{R}^{d_\text{out} \times d_\text{in}}$, with $d_\text{in}$ the input and $d_\text{out}$ the output dimension.

\subsection{Continual Learning}
We assume the initial model, with parameters $\bm\theta^0$, has been trained on an initial task set $\T_0$. The model is adapted sequentially to a stream of new tasks $\T_1,\ldots,\T_t$. At each stage $i>0$, adaptation starts from the parameters $\bm\theta^{i-1}$ obtained after the previous task and produces updated parameters $\bm\theta^i$. The objective is to perform well on the current task $\T_i$, but also to preserve performance on all previously seen tasks $\T_0,\ldots,\T_{i-1}$. Each task $\T_i$ is associated with a dataset $(\bm X, \bm y) \in \D_i$ of paired speech. In the CL setting studied here, data from previous tasks, $\D_0,\ldots,\D_{i-1}$, is no longer available when adapting to $\T_i$, or is too costly to reintroduce at scale. This is particularly relevant for foundation models, whose initial training involves hundreds of thousands of hours of speech \cite{owsmv4}. The central challenge is therefore to acquire new task-specific knowledge while avoiding catastrophic forgetting of the capabilities present in the initial model.

\subsection{Parameter Efficient Fine-Tuning}
\label{subsec:pecl}
In addition to avoiding forgetting, adaptation should be memory- and compute-efficient. Since foundation models are generally large, fully fine-tuning all parameters is costly. Parameter-efficient methods aim to reduce this overhead by adapting only a small number of additional parameters. Most parameter-efficient fine-tuning methods focus on the linear weight matrices $\bm W \in \mathbb{R}^{d_\text{out} \times d_\text{in}}$ of the model and adapt them while keeping the remaining weights frozen. 

Low-Rank Adaptation (LoRA)~\cite{lora} trains a low-rank update $\Delta\bm W=\bm B\bm A$, with $\bm B \in \mathbb{R}^{d_\text{out}\times r}$ and $\bm A \in \mathbb{R}^{r \times d_\text{in}}$, where $r \ll \min(d_\text{out}, d_\text{in})$, and adds this update to $\bm W$. Another example is Structured SVD (SSVD)~\cite{wang2025ssvd}, which considers the SVD of $\bm W$ and introduces rescaling and rotation matrices operating on the top-$k$ singular directions, while freezing the singular values and vectors. 

\subsection{Gradient Low-Rank Projection with GaLore}
\label{subsec:galore}
Gradient Low-Rank Projection (GaLore)~\cite{galore} builds on the observation that, although a weight matrix $\bm W \in \mathbb{R}^{d_\text{out} \times d_\text{in}}$ need not be low-rank, its gradients often exhibit low-rank structure. GaLore exploits this property by applying the optimizer in a low-dimensional gradient subspace, thereby reducing the memory required by optimizer states such as those used by Adam~\cite{adam}, AdamW~\cite{adamw}, and Adagrad~\cite{adagrad}, whose states typically scale with the number of parameters and can exceed the memory needed to store the parameters themselves \cite{galore}.

Let $\bm G_s \in \mathbb{R}^{d_\text{out} \times d_\text{in}}$ denote the gradient of weight matrix $\bm W$ at optimization step $s$:
\begin{equation}
\bm G_s =
-\frac{\partial \L(\bm X_s, \bm y_s; \bm \theta_{s-1})}
{\partial \bm W}.
\label{eq:grad}
\end{equation}
where $(\bm X_s, \bm y_s)$ is the paired sample or mini-batch at step $s$ and $\bm \theta_{s-1}$ are the model parameters from step $s-1$.

Given orthogonal matrices $\bm P \in \mathbb{R}^{d_\text{out} \times r}$ and $\bm R \in \mathbb{R}^{d_\text{in} \times r}$, with $r \ll \min(d_\text{out},d_\text{in})$, GaLore projects the gradient into an $r \times r$ optimizer space:
\begin{equation} \tilde{\bm G_s} = \bm P^\top \bm G_s \bm R \in \mathbb{R}^{r \times r} 
\label{eq:galore_proj} 
\end{equation}
The optimizer is applied to $\tilde{\bm G}_s$, i.e., $\tilde{\bm G}^{\mathrm{OPT}}_s=\text{Optimizer}(\tilde{\bm G}_s)$, so that its states are stored in the $r \times r$ projected space rather than in the original $d_\text{out} \times d_\text{in}$ space. The resulting optimizer update $\tilde{\bm G}^{\mathrm{OPT}}$  is then mapped back to the original parameter space:
\begin{equation}
\bm G^{\mathrm{OPT}}_s
=
\bm P \tilde{\bm G}^{\mathrm{OPT}}_s \bm R^\top,
\label{eq:back_to_original_dimensions}
\end{equation}
and used to update the weight matrix $\bm W_{s-1}$ from $\bm \theta_{s-1}$ to $\bm W_s$:
\begin{equation}
\bm W_s = \bm W_{s-1} + \eta \bm G_s^{\mathrm{OPT}},
\label{eq:galore_step}
\end{equation}
with $\eta$ the learning rate. Thus, unlike PEFT methods, GaLore does not reduce the number of trainable parameters, nor is its adaptation low-rank; instead, it achieves memory-efficient full-parameter adaptation by reducing the dimensionality of the optimizer states.

To construct the projection matrices, GaLore computes the singular value decomposition of the current gradient,
\begin{equation}
\bm G_s
=
\bm U_s \bm \Sigma_s \bm V_s^\top
=
\sum_{j=1}^{k}
\sigma_{s,j} \bm u_{s,j} \bm v_{s,j}^{\top},
\label{eq:compute_svr_galore}
\end{equation}
where $k=\min(d_\text{out},d_\text{in})$. The projection matrices are chosen as the leading singular vectors, 
\begin{equation}
\bm P=\bm U_{s,:r},
\qquad
\bm R=\bm V_{s,:r}.
\label{eq:set_pr_galore}
\end{equation}
Since recomputing the SVD at every step is costly (and suboptimal \cite{galore}), GaLore keeps $\bm P$ and $\bm R$ fixed for $T$ optimization steps and updates them using the SVD of gradient $\bm G_{s+T}$ after this interval.

\section{Curvature-Guided GaLore}

GaLore reduces optimizer-state memory, but its projection bases are selected from the current gradient alone and provide no protection against forgetting. We propose Curvature-Guided GaLore (CGaLore), which uses old-task curvature to guide the construction of the low-rank optimizer subspace. CGaLore is summarized in Algorithm \ref{alg:cgalore}.

\subsection{Integrating Old-Task Curvature Information}
\label{subsec:integrating}
Let $S$ denote some data from the initial task set $\T_0$, i.e., $S \subseteq \D_0$, and define the old-task loss as $\L_0(\bm\theta)=\frac{1}{|S|}\sum_{(\bm X,\bm y)\in S}\L(\bm X,\bm y;\bm\theta)$. Since $\bm\theta^0$ was optimized on $\D_0$, we assume that $\nabla_{\bm \theta} \L_0(\bm\theta^0)\approx \bm 0$. A second-order Taylor expansion around $\bm\theta^0$ then gives
\begin{equation}
    \L_0(\bm\theta^0+\Delta\bm\theta)
    \approx
    \L_0(\bm\theta^0)
    +
    \frac{1}{2}
    \Delta\bm\theta^\top
    \mathcal{\bm B}_0
    \Delta\bm\theta,
\end{equation}
where $\mathcal{\bm B}_0=\nabla^2_{\bm\theta}\L_0(\bm\theta^0)\in \mathbb{R}^{N\times N}$
is the Hessian of the old-task loss at the initial model.

Let $\bm g = -\nabla_{\bm\theta} \L(\bm X, \bm y; \bm \theta^0) \in \mathbb{R}^{N}$ be the descent direction for a sample $(\bm X,\bm y)\in \D_1$ from a new task $\T_1$. Updating the model directly along $\bm g$ may cause forgetting on $\T_0$. We therefore consider the direction that maximizes alignment with $\bm g$ while keeping the quadratic increase in old-task loss bounded:
\begin{equation}
\max_{\bm p} \ \bm p^\top \bm g
\quad \text{s.t.} \quad
\frac{1}{2}\bm p^\top \mathcal{\bm B}_0 \bm p \leq \tau .
\end{equation}
The solution is proportional to the inverse-Hessian direction, i.e. $\bm p \propto \mathcal{\bm B}_0^{-1}\bm g$: multiplying the new-task gradient by the inverse old-task Hessian yields a direction aligned with the new task that limits the quadratic increase in old-task loss. As CGaLore uses this result only to identify directions that are useful for the new task and safe for the old tasks, the proportionality constant $\tau$ is irrelevant.

In practice, forming the Hessian $\mathcal{\bm B}_0$ is infeasible. Prior work approximates the Hessian with Kronecker-factored approximate curvature (KFAC)~\cite{kf} and applies the inverse-Hessian correction at layer level~\cite{vanderEeckt2026inversehessian}. Importantly, \cite{vanderEeckt2026inversehessian} applies the correction post hoc to the final parameter displacement after fine-tuning. In contrast, we use KFAC during optimization to guide the projection bases.

\subsection{Curvature-Guided Subspace Selection}
\label{subsec:cgalore}

Consider $\bm W^0$ the parameters of the weight matrix $\bm W$ from the initial model parameters $\bm \theta^0$. The Hessian $\bm B$ of the loss on old tasks $\T_0$ around $\bm W^0$, approximated using KFAC, is given by:
\begin{equation}
    \bm B = \frac{\partial^2 \L_0(\bm \theta)}{\partial  (\bm W^0)^2} \approx \bm Q \otimes \bm H
    \label{eq:q_h}
\end{equation}
with $\bm Q \in \mathbb{R}^{d_\text{in} \times d_\text{in}}$ the input covariance and $\bm H \in \mathbb{R}^{d_\text{out} \times d_\text{out}}$ output-gradient covariance, so that $\bm Q \otimes \bm H$ is positive-definite. 

To update the orthogonal bases $\bm P$ and $\bm R$ every $T$ steps, CGaLore starts from the current gradient $\bm G_s$ and applies the inverse old-task curvature estimate from Eq.~\eqref{eq:q_h}, following Sec. \ref{subsec:integrating}, before computing the SVD. At the layer level, using the KFAC, this gives
\begin{equation}
    \bm G^{\text{KFAC}}_s = \bm H^{-1} \bm G_s \bm Q^{-1}
    \label{eq:kfac_proj}
\end{equation}
Next, as in Eq. \eqref{eq:compute_svr_galore}, we apply SVD to the resulting gradient:
\begin{equation}
    \bm G^{\text{KFAC}}_s = \bm U^{\text{KFAC}}_s \bm \Sigma^\text{KFAC}_s (\bm V^\text{KFAC}_s)^\top
    \label{eq:kfac_svd}
\end{equation}
and, following Eq. \eqref{eq:set_pr_galore}, find $\bm P$ and $\bm R$ by keeping only the $r$ singular directions associated with highest singular values. The resulting projection matrices $\bm P$ and $\bm R$ therefore span directions that are both relevant for the new task, because they are derived from the current-task gradient, and less likely to interfere with old tasks, because the gradient is filtered by the inverse old-task curvature before the singular directions are selected.

For any gradient $\bm G_p$ with $p>s$, we now proceed as in
Eqs.~\eqref{eq:galore_proj}--\eqref{eq:galore_step} from GaLore. As in GaLore, moreover, we update $\bm P, \bm R$ every $T$ iterations using Eqs. ~\eqref{eq:kfac_proj}, \eqref{eq:kfac_svd} and \eqref{eq:set_pr_galore}.

\subsection{Two-Sided Projection}
\label{subsec:two_sided}
While GaLore can project only on the left or right side of $\bm G_s$ in Eq. \eqref{eq:galore_proj}, CGaLore uses the full projection. This allows the update subspace to be restricted in both the output and input dimensions of $\bm W$. Since CGaLore uses these bases not only for memory reduction but also to bias updates toward directions with low old-task curvature, constraining both sides gives a more complete layer-wise control over the update than using only $\bm P$ or only $\bm R$.

\subsection{Maintaining KFAC statistics}
Foundation models are often trained on hundreds of thousands of hours of speech~\cite{owsmv4}, making it impractical to compute KFAC statistics on all pretraining data. For the initial adaptation, as described in Sec.~\ref{subsec:integrating}, we therefore sample a subset $S \subseteq \D_0$ and compute the KFAC factors $\bm Q$ and $\bm H$ on the initial model $\bm\theta^0$, yielding the layer-wise approximation $\bm B \approx \bm Q \otimes \bm H$ for each weight matrix $\bm W^0$.

After adapting to a subsequent task $\T_i$, we sample a subset $S_i \subseteq \D_i$ and compute task-specific KFAC factors $\tilde{\bm Q}_i$ and $\tilde{\bm H}_i$ using the adapted model. We then update the stored KFAC factors by averaging:
\begin{equation}
\bm Q \gets (1-\alpha)\bm Q+\alpha \tilde{\bm Q}_i,
\qquad
\bm H \gets (1-\alpha)\bm H+\alpha \tilde{\bm H}_i,
\label{eq:update_kfac_factors}
\end{equation}
with $0 \leq \alpha \leq 1$. This provides a compact approximation of the accumulated old-task curvature: instead of storing separate KFAC factors for every previous task, CGaLore keeps only one running set of factors per layer, which can later be used to adapt to task $\T_{i+1}$.

\subsection{Implementation and Efficiency}
CGaLore uses KFAC only when updating the GaLore projection bases. The KFAC factors are precomputed from old-task data and stored in CPU memory; at an update step, the factors for the current layer are temporarily moved to the GPU to compute Eq. \eqref{eq:kfac_proj}. Thus, KFAC does not increase the GPU memory footprint, but only adds computation during the subspace update. To reduce the update cost, we use randomized SVD~\cite{halko2011finding} instead of exact SVD, following~\cite{galore2}.

\begin{algorithm}
\caption{Curvature-Guided GaLore (CGaLore)}
\label{alg:cgalore}
\begin{algorithmic}[1]
\Require Previous model $\bm{\theta}^{i-1}$, data $\mathcal{D}_i$ for new task $\T_i$, KFAC factors ${\bm Q,\bm H}$, rank $r$, update interval $T$
\State Set $s\gets 1$, $\bm \theta_{s-1} \gets \bm \theta^{i-1}$
\For{each optimization step $s$ on $(\bm X, \bm y) \in \mathcal{D}_i$}
\For{each linear layer with weights $\bm W_{s-1} \in \bm \theta_{s-1}$}
\State Compute gradients $\bm G_s$ using Eq. \eqref{eq:grad}.
\If{$s=1$ or $s \bmod T = 1$}
\State Update $\bm P, \bm R$ through Eqs. \eqref{eq:kfac_proj}, \eqref{eq:kfac_svd} and \eqref{eq:set_pr_galore}.
\EndIf
\State Update $\bm W_{s-1}$ into $\bm W_s$ using Eqs. \eqref{eq:galore_proj}, \eqref{eq:back_to_original_dimensions} and \eqref{eq:galore_step}.
\State Store $\bm W_s$ in $\bm \theta_s$
\EndFor
\EndFor
\State Update KFAC factors $\bm Q, \bm H$ using data $S_i \subseteq \D_i$ and Eq. \eqref{eq:update_kfac_factors}.
\State \Return Adapted model parameters $\bm{\theta}^{i}=\bm \theta_s$
\end{algorithmic}
\end{algorithm}

\section{Experiments}

\begin{table*}
\centering
\begin{threeparttable}
\caption{Results of the experiments. ``Method'' denotes the base adaptation method, while ``CL'' denotes the continual learning method applied on top of it. WERs are measured after learning all tasks. Negative BWT indicates forgetting.}
\setlength{\tabcolsep}{4pt}
\begin{tabular}{
l l c@{\hspace{5pt}} c@{\hspace{5pt}} c@{\hspace{5pt}} c@{\hspace{5pt}} c@{\hspace{5pt}} c@{\hspace{5pt}} c@{\hspace{5pt}}
c@{\hspace{5pt}} c@{\hspace{5pt}} c@{\hspace{5pt}} c@{\hspace{5pt}} c@{\hspace{5pt}} c@{\hspace{5pt}} c@{\hspace{5pt}}
}
\toprule
& & \multicolumn{7}{c}{\textbf{\textit{Experiment 1}}} & \multicolumn{7}{c}{\textbf{\textit{Experiment 2}}} \\
\cmidrule(lr){3-9} \cmidrule(lr){10-16}
& & \multicolumn{5}{c}{\textbf{WER$\downarrow$ per task}} & \multicolumn{2}{c}{\textbf{Average}} & \multicolumn{5}{c}{\textbf{WER$\downarrow$ per task}} & \multicolumn{2}{c}{\textbf{Average}} \\
\cmidrule(lr){3-7} \cmidrule(lr){8-9} \cmidrule(lr){10-14} \cmidrule(lr){15-16}
\textbf{Method} & \textbf{CL} &
\textbf{1--ENG} & \textbf{1--DEU} & \textbf{1--ESP} &
 \textbf{2--L2/1} & \textbf{3--L2/2} &
\textbf{WER$\downarrow$} & \textbf{BWT$\uparrow$} & \textbf{1--ENG} & \textbf{1--DEU} & \textbf{1--ESP} &
 \textbf{2--CNL} & \textbf{3-CVL} &
\textbf{WER$\downarrow$} & \textbf{BWT$\uparrow$}   \\
\midrule
Initial model & --  & 13.4 & 11.3 & 11.3 & 8.2 & 16.7 & 12.18 & -- & 
13.4 & 11.3 & 11.3 & 45.7 & 37.9 & 22.48  & --  \\
\subrule
\multirow{4}{*}{Full Fine-Tuning} & -- & 
21.0 & 17.1 & 15.9 & 1.7 & 2.3 & 11.60 & \phantom{1}-4.4 &
24.3 & 57.3 & 21.7 & 27.7 & 13.6 & 28.94 & -18.2 \\
& IHR & 
13.4 & 11.2 & 11.3 & 3.3 & 7.9 & \phantom{1}9.39 & \phantom{-1}0.3 &
 13.7 & 11.6 & 11.6 & 31.8 & 23.4 & 18.41 & \phantom{-1}1.0\\
& SVR & 15.2 & 12.2 & 11.8 & 1.6 & 3.6 & \phantom{1}8.88 & \phantom{1}-0.5 & 16.9 & 22.7 & 15.0 & 27.5 & 15.6 & 19.52 & \phantom{1}-4.6 \\
 & Sep. Model  &
13.4 & 11.3 & 11.3 & 2.1 & 2.3 & \phantom{1}8.07 & \phantom{-1}0.0 &
13.4 & 11.3 & 11.3 & 22.4 & 13.6 & 14.38 & \phantom{-3}0.0  \\
\midrule
\multirow{2}{*}{LoRA} & -- &
25.1 &  31.2 & 27.9 & 2.2 & 2.6 & 17.79 & -11.9 &
41.8 & 88.4 & 44.5 & 28.9 & 14.7 & 43.66 & -35.7  \\
 & FTA &
14.4 & 12.8 & 12.4 & 3.0 & 7.5 & 10.01 & \phantom{1}-0.6 &
16.4 & 20.6 & 15.1 & {27.2} & 19.0 & 19.64 & \phantom{1}-3.6  \\
\subrule
GaLore & -- & 20.8 & 17.6 & 16.9 & 1.8 & 2.7 & 11.98 & \phantom{1}-4.7  &
23.7 & 53.1 & 22.2 & 30.6 & 15.1 & 28.93  & -17.2 \\
\subrule 
BiLoRA & -- &
18.1 & 14.7 & 14.3 & 1.7 & 3.1 & 10.37 & \phantom{1}-2.5 &
19.4 & 34.7 & 17.8 & 30.0 & 16.0 & 23.58  & \phantom{1}-9.7  \\
\subrule 
CSSVD & -- & 
14.1 & 11.9 & 12.0 & 2.8 & 6.8 & \phantom{1}9.51 & \phantom{1}-0.3 &
{14.9} & {14.4} & {13.4} & {28.6} & 20.4 & {18.33} & \phantom{1}{-1.9}  \\
\midrule
CGaLore & -- & 15.1 & 11.5 & 11.6 & 2.0 & 3.5 & \phantom{1}\textbf{8.72}\tnote{a} & \phantom{1}-0.3 & 
15.5 & 14.5 & 13.0 & 28.9 & 16.4 & \textbf{17.68}\tnote{a} & \phantom{1}-2.7 \\
\bottomrule
\end{tabular}
\begin{tablenotes}
\footnotesize
\item[a] Significantly outperforms all baselines except the full fine-tuning separate-model reference, which serves as an upper bound.
\end{tablenotes}
\label{tab:results}
\end{threeparttable}
\end{table*}
Experiments are done in ESPnet2 \cite{watanabe2018espnet}. Code, configurations, and data splits are available at \href{https://github.com/StevenVdEeckt/cgalore}{github.com/StevenVdEeckt/cgalore}.


\noindent \textbf{Model.} The model is the Open Whisper-style Speech Model (OWSM) v3.2 small \cite{owsm}, consisting of 9 E-Branchformer \cite{e_branchformer} encoder and 9 Transformer \cite{transformer} decoder layers. The model, with a vocabulary of $50{,}000$ output tokens, containing 366.7M parameters, was trained on 180k hours of speech from 151 languages. For adaptation, only the weight matrices $\bm W$ of the linear layers are updated, while output layers are frozen. We integrate (C)GaLore with AdamW \cite{adamw} (using learning rate $5e\text{-}4$) without scheduler, training the model for 20 epochs with an effective batch size of 64. 

\noindent\textbf{Data.} We conduct two experiments. For both, the initial tasks $\T_0$ are English (ENG), German (DEU), and Spanish (ESP) from Common Voice \cite{commonvoice}, used to measure forgetting. In Exp.~1, we adapt OWSM to L2-Arctic \cite{zhao2018l2arctic}, which contains English speech from non-English speakers. We split L2-Arctic into two tasks based on accent, referred to as L2/1 and L2/2. The second experiment resembles \cite{svr}, using Corpus Gesproken Nederlands (CGN) \cite{cgn}, split into Dutch from the Netherlands (CNL) and Belgium (CVL), yielding two tasks. CGN includes varied Dutch speech from settings such as interviews, lectures, and broadcasts.

\noindent\textbf{Baselines}. 
We compare against the following methods and describe how each adapts linear layer $\bm W \in \mathbb{R}^{d_\text{out} \times d_\text{in}}$ to the new task:
\begin{enumerate}[label=(\alph*)]
    \item LoRA \cite{lora}: learns a low-rank adaptation as discussed in Sec. \ref{subsec:pecl}, without tackling catastrophic forgetting.
    \item LoRA + FTA \cite{cssvd, ugan25_interspeech}: combines LoRA with Fine-Tuning with Averaging (FTA) \cite{weight_averaging}, averaging the low-rank updates $\Delta \bm W$ with the pretrained weight matrix $\bm W$.    
    \item GaLore \cite{galore}, discussed in Sec. \ref{subsec:galore}.
    \item BiLoRA \cite{bilora}: uses a bilinear update in fixed orthogonal bases, i.e., $\Delta \bm W= \bm F_{\text{out}}\,\bm B\,\bm F_{\text{in}}^{H}$, with $\bm F_{\text{out}}$ and $\bm F_{\text{in}}$ the 1D discrete Fourier transform matrices so that only $k_B$ elements in $\bm B \in \mathbb{C}^{d_\text{out}\times d_\text{in}}$ are trained. 
    \item Continual SSVD (CSSVD) \cite{cssvd}: builds on SSVD \cite{wang2025ssvd}, explained in Sec. \ref{subsec:pecl}, instead learning only rotations between the bottom-$k$ singular directions on the new task. 
    \item Full Fine-Tuning (FFT): we also fully fine-tune the linear weight matrices, which will result in forgetting. As additional baselines, we combine FFT with (a) Inverse Hessian Regularization (IHR) \cite{vanderEeckt2026inversehessian}, which applies the inverse curvature correction post-hoc to the update $\Delta \bm W$; (b) Singular Value Rehearsal \cite{svr}, which applies SVD to the update $\Delta \bm W$ and learns on a small memory buffer of old data which directions to keep; (c) Sep. Model, which provides a best-case scenario, keeping a separate model $\bm \theta^i$ per task $\T_i$ using a task oracle - it does therefore not forget. SVR is a rehearsal-based method, requiring storage of past data. 
\end{enumerate}
PECL methods learn 9.0M, and FFT and (C)GaLore 244.8M parameters; with $r=20$ for LoRA-based methods and $p=0.40$ for CSSVD. For SVR, as in \cite{svr}, we store 50 utterances (approx. 4 minutes) in the memory buffer, divided equally across old tasks. 

\noindent \textbf{Implementation.} For (C)GaLore, we set $r=215$, the optimizer states thus having approximately the same memory budget as PECL baselines, and update the projection bases every $T=200$ optimization steps. For CGaLore, KFAC factors are estimated using all three old tasks, with 20k utterances per task, corresponding to 28 hours of speech per task, and $\alpha=1/4$. We use KFAC damping of $10^{-3}$.

\noindent \textbf{Metrics.} 
We report word error rate (WER, \%) for the final model on each task, and use Average WER over all learned tasks as main metric. Forgetting is measured by backward transfer (BWT), defined as: $\text{BWT}=1/i \sum_{j=0}^{i-1} (\text{WER}_j^j - \text{WER}_j^i)$, where $\text{WER}_k^l$ represents WER on task $\T_k$ after learning task $\T_l$, and $\T_i$ is the most recently learned task. In our setting, $\T_0$ contains multiple tasks (Sec. \ref{subsec:notation}), which are treated individually when computing BWT. Negative values indicate mean WER increase and degradation on old tasks. Average WER significance is assessed with a Wilcoxon signed-rank test on per-utterance error counts~\cite{Strik2000ComparingTR} at 0.1\% level.

\section{Results}
\label{sec:results}


Table \ref{tab:results} shows the results of the two experiments. 

\subsection{Experiment 1: L2-Arctic}
\label{sec:results_l2arctic}
Exp.~1 finds all methods substantially improving performance on  L2-Arctic compared with the initial model. LoRA and GaLore reach new-task WERs close to full fine-tuning, but differ strongly in forgetting: LoRA suffers much larger degradation on the initial tasks, whereas GaLore reduces this effect but still obtains a BWT of -4.7. 

CGaLore gives the best balance between learning and retention. It achieves the lowest forgetting among parameter- and memory-efficient methods, with BWT of -0.3, while also learning the new tasks better than LoRA+FTA and CSSVD. As a result, CGaLore obtains the best Average WER of 8.72, improving over CSSVD, the strongest baseline, by 8.3\%. Compared with GaLore, CGaLore eliminates 93.6\% of forgetting while recovering 94.3\% of the L2/2 performance gap, closing 83.4\% of the gap between GaLore and the Full Fine-Tuning + Sep. Model reference. CGaLore even outperforms all FFT baselines, including SVR which relies on storage of past data.

\subsection{Experiment 2: CGN}
\label{sec:results_cgn}
Similar conclusions hold for Exp.~2, although forgetting is on average substantially higher for most methods. LoRA and GaLore again approach the new-task performance of full fine-tuning, but both suffer strong degradation on the initial tasks.

CGaLore again achieves the best average WER, with 17.68. It provides the strongest balance between learning and retention: compared with CSSVD, it has slightly higher forgetting but better new-task performance. Compared with GaLore, CGaLore reduces forgetting by 84.3\% while retaining 94.3\% of the CVL performance, and closes 77.4\% of the gap between GaLore and the FFT separate-model reference. CGaLore also outperforms all FFT baselines

\subsection{Effect of KFAC Estimation Data}
\label{sec:kfac_data}
Table~\ref{tab:heldout_forgetting} reports WERR (WER reduction) on old tasks after the first adaptation of Exp.~1 and Exp.~2. Rows indicate which old-task data are used to compute KFAC. We report WERR both for the main old tasks used in the primary evaluation (ENG, DEU, ESP) and for held-out old tasks not used for KFAC: TEDLIUM~\cite{tedlium} (TED), Swedish (SWE), Italian (ITA), and Russian (RUS) from Common Voice. Negative WERR indicates forgetting.

Moving from GaLore, which uses no KFAC, to CGaLore substantially reduces forgetting on both main and held-out old tasks. In Exp.~1, using ENG for KFAC further reduces forgetting on ENG, while the effect of including DEU is small. For ESP, using only ENG appears to bias the update toward directions safe for ENG and DEU but less safe for ESP. The held-out tasks generally follow their closest main-task counterpart: ITA follows ESP, SWE behaves similarly to DEU, and RUS benefits most from the more diverse KFAC estimate.

In Exp.~2, including DEU in the KFAC estimate is especially important. Since the new Dutch task is close to German, interference with DEU is high, as also reflected by the strong forgetting of FFT, LoRA, and GaLore in Table~\ref{tab:results}. Using DEU for KFAC therefore greatly reduces forgetting on DEU. SWE, another related language, follows the same trend, although the reduction is less pronounced; using SWE additionally for KFAC might further reduce forgetting.

These results show that CGaLore also reduces forgetting on held-out old tasks that were not used for KFAC estimation. This is important for foundation-model adaptation: CGaLore does not require access to all tasks used during pretraining to provide protection against forgetting. Nevertheless, KFAC is most effective when it includes old tasks that strongly interfere with the new task.

\begin{figure}
    \centering
    \includegraphics[width=\linewidth]{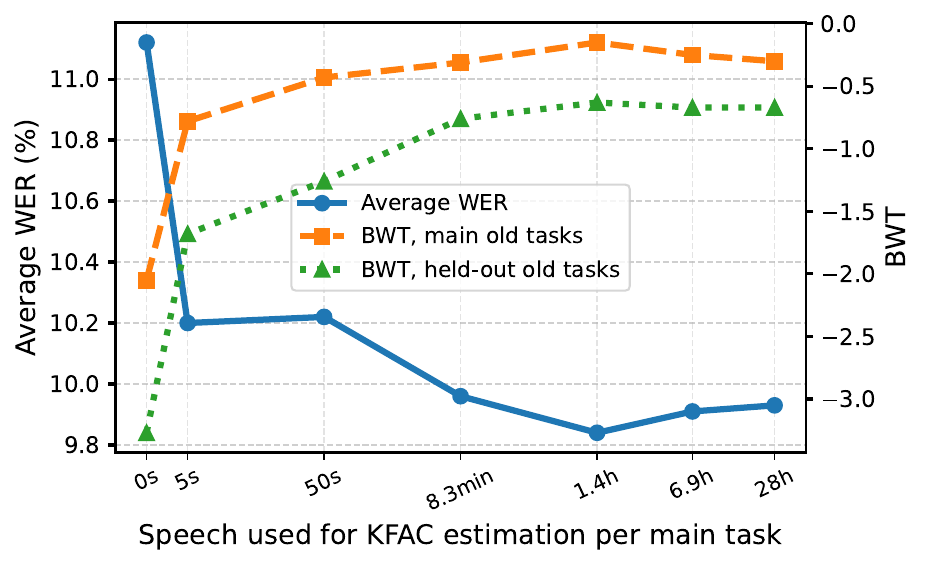}
    \caption{Performance of CGaLore as a function of speech used (per task) for KFAC estimation the first adaptation of Exp. 1. ENG, DEU and ESP are used for KFAC estimation. The blue curve shows the Average WER (on ENG, DEU, ESP and L2/1), orange the BWT on ENG, DEU and ESP; and green the BWT on TED, SWE, ITA and RUS. Negative BWT indicates forgetting. }
    \label{fig:exp_kfac_amount_of_data}
\end{figure}

Fig. ~\ref{fig:exp_kfac_amount_of_data} further studies how much data is needed to estimate KFAC. We vary the amount of speech per main old task from 0 seconds, corresponding to GaLore without KFAC, to 28 hours, using ENG, DEU, and ESP for KFAC estimation. We report average WER, BWT on the main old tasks, and BWT on the held-out old tasks.

Reducing the KFAC data from 28 hours to 8.3 minutes per task has little effect: average WER and BWT remain nearly unchanged. Performance starts to degrade only when reducing the data to 50 seconds per task, and forgetting increases further at 5 seconds per task. Even then, CGaLore still reduces GaLore's forgetting by more than 60$\%$. Thus, while more KFAC data is beneficial, a small amount of old-task speech is sufficient for CGaLore to reduce forgetting.

\begin{table}
    \centering
    \caption{
    WERR (WER reduction, $\%$) on old tasks after adaptation.
    Rows indicate the old-task data used to compute KFAC; ``--'' denotes GaLore without KFAC.
    Negative WERR indicates forgetting.
    WERR is given for main (ENG, DEU, ESP) and held-out old tasks not used for KFAC: TEDLIUM (TED)~\cite{tedlium}, Swedish (SWE), Italian (ITA) and Russian (RUS) from Common Voice.
    }
    \begin{threeparttable}
        \label{tab:heldout_forgetting}
    \begin{tabular}{l c@{\hspace{5pt}} c@{\hspace{5pt}} c@{\hspace{5pt}} c@{\hspace{5pt}} c@{\hspace{5pt}} c@{\hspace{5pt}} c@{\hspace{5pt}}}
    \toprule
    \multicolumn{1}{c}{\multirow{2}{*}{\textbf{Tasks for KFAC}}} & 
    \multicolumn{3}{c}{\textbf{Per main task}}
    & \multicolumn{4}{c}{\textbf{Per held-out task}}   \\
    \cmidrule(lr){2-4} \cmidrule(lr){5-8}
    & \textbf{ENG} & \textbf{DEU} & \textbf{ESP}
    & \textbf{TED} 
    & \textbf{SWE} 
    & \textbf{ITA} 
    & \textbf{RUS} \\
    \midrule
    \multicolumn{8}{c}{\textit{Experiment 1 -- Adaptation to 2--L2/1}} \\
    \cmidrule(lr){1-8}
    -- & -18.7 & -16.8 & -15.0 & \phantom{1}-7.0 & -15.0 & -17.7 & -13.7 \\
    ENG+DEU+ESP & \phantom{1}-3.7 & \phantom{1}-0.9 & \phantom{1}-2.7 & \phantom{1}-4.1 & \phantom{1}-3.2 & \phantom{1}-3.2 & \phantom{1}-1.0   \\
    ENG & \phantom{1}-0.7 & \phantom{1}-0.9 & \phantom{1}-6.2  &  \phantom{1}-2.2 & \phantom{1}-4.5 & \phantom{1}-5.5 & -\phantom{1}4.0  \\
    DEU & \phantom{1}-6.7 & \phantom{1}-1.8 & \phantom{1}-2.7 &   \phantom{1}-4.5 & \phantom{1}-3.9 & \phantom{1}-2.9 & \phantom{1}-3.5  \\
    \midrule
    \multicolumn{8}{c}{\textit{Experiment 2 -- Adaptation to 2--CNL}} \\
    \cmidrule(lr){1-8}
    -- & -38.1 & -134.5 & -46.0 & -34.0 & -51.7 & -56.3 & -60.5 \\
    ENG+DEU+ESP & -10.4 & \phantom{1}-15.9  & \phantom{1}-6.6  & -14.4 & -23.5 & \phantom{1}-9.8  & \phantom{1}-6.7    \\
    ENG & \phantom{1}-9.7  & \phantom{1}-40.7 & -13.3 &  -12.0 & -28.9 & -15.7 & -12.3   \\
    DEU & -14.2 & \phantom{11}-9.7  & -14.2 & -16.2 & -22.6 & -11.0 & \phantom{1}-6.0     \\
    \bottomrule    
    \end{tabular}
    \end{threeparttable}
\end{table}

\subsection{Memory and Runtime Efficiency}
\label{sec:memory_efficiency}
Table~\ref{tab:efficiency} studies the training efficiency of CGaLore on the first adaptation of Exp.~1. GaLore and CGaLore do not increase peak GPU memory relative to LoRA: both use 10.56~GB, slightly below LoRA at 10.90~GB, despite updating full weight matrices rather than only LoRA parameters. Although GaLore makes the optimizer step substantially more expensive, with optimizer time 12.6$\times$ higher than LoRA, its unchanged forward/backward graph makes total training approximately 11$\%$ faster overall in our setting with gradient accumulation of 16. CGaLore adds only modest overhead over GaLore: KFAC multiplication accounts for 19$\%$ of the randomized-SVD time, increasing optimizer time by 1.3$\%$ and total training time by only 0.4$\%$. Thus, CGaLore retains GaLore's practical efficiency while adding curvature-guided subspace selection.

\begin{table}[t]
\centering
\caption{Training efficiency comparison on first adaptation of Exp.~1, on a single NVIDIA A100 GPU. Peak GPU memory denotes peak allocated memory. Relative time is normalized to LoRA.}
\label{tab:efficiency}
\begin{tabular}{lccc}
\toprule
\multirow{2}{*}{\textbf{Method}} & \textbf{Peak GPU} & \textbf{Optimizer time} & \textbf{Train time} \\
\cmidrule(lr){2-2} \cmidrule(lr){3-3} \cmidrule(lr){4-4}
       & \textbf{GB $\downarrow$} & \textbf{relative $\downarrow$} & \textbf{relative $\downarrow$} \\
\midrule
LoRA    & 10.90 & \phantom{3}1.0$\times$ & 1.00$\times$   \\
GaLore  & 10.56 & 12.6$\times$ & 0.89$\times$   \\
CGaLore & 10.56 & 12.8$\times$ & 0.90$\times$ \\
\bottomrule
\end{tabular}
\end{table}

\subsection{Effect of Subspace Update Interval}
\label{sec:adaptive_refresh}
\begin{figure}
    \centering
    \includegraphics[width=\linewidth]{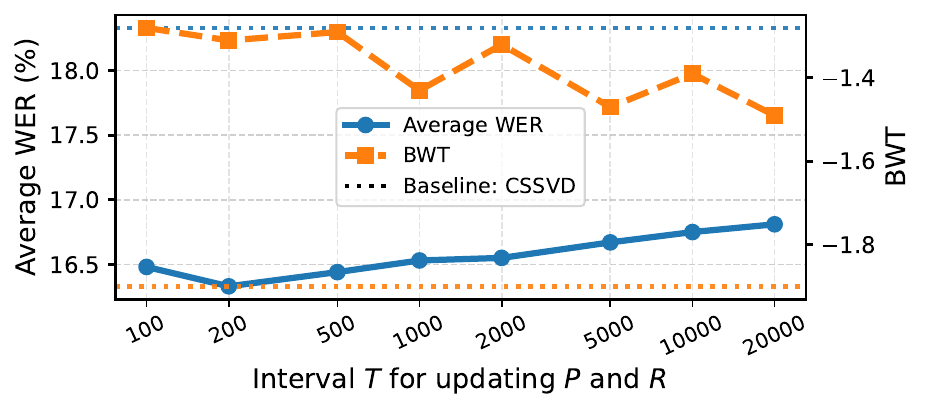}
    \caption{Average WER and BWT as a function of the interval $T$ in CGaLore on first adaptation of Exp.~2.
$T$ determines how often $\bm P$ and $\bm R$ are updated.}
    \label{fig:adaptive_refresh_tradeoff}
\end{figure}

Fig.~\ref{fig:adaptive_refresh_tradeoff} studies the effect of the projection update interval $T$ on average WER and BWT for the first adaptation of Exp.~2. Performance is best at $T=200$, decreases only slightly up to $T=2000$, and degrades more noticeably for larger intervals. This degradation mainly affects new-task performance, since BWT remains relatively flat across $T$. Thus, CGaLore is not highly sensitive to the exact update interval, consistent with observations for GaLore~\cite{galore}; even at $T=20000$, the gap to CSSVD remains substantial.

\subsection{Interaction with IHR}
\label{subsec:interaction_with_ihr}

\begin{table}
    \centering
    \caption{Comparing GaLore and CGaLore with and without post-hoc IHR correction on the first (2--CNL) and second (3--CVL) adaptation of Exp. 2. $\lambda$ is the hyper-parameter controlling the rotation of the update, as discussed in Sec. \ref{subsec:interaction_with_ihr}.}
    \begin{threeparttable}
    \begin{tabular}{l l l c c c c}
    \toprule
    \multirow{2}{*}{\textbf{Method}} & \multirow{2}{*}{\textbf{IHR}} & \multirow{2}{*}{\textbf{$\lambda$}} &  \multicolumn{2}{c}{\textbf{2--CNL: Average}} & \multicolumn{2}{c}{\textbf{3--CVL: Average}} \\
    \cmidrule(lr){4-5} \cmidrule(lr){6-7}
     & & & \textbf{WER}$\downarrow$ & \textbf{BWT}$\uparrow$ & \textbf{WER}$\downarrow$ & \textbf{BWT}$\uparrow$  \\
    \toprule 
    GaLore & \xmark & -- &  21.56 & -8.5 & 28.93 & -17.2 \\
    CGaLore & \xmark & -- &  16.33\tnote{a} & -1.3 & 17.68 & \phantom{1}-2.7 \\
    \midrule
    GaLore & \cmark & 0.50 &  16.58\tnote{a} & -1.2 & 17.74\tnote{a}  & \phantom{1}-2.3 \\
    CGaLore & \cmark & 0.75 & 16.09\tnote{a} & -0.6 & 16.86\tnote{a} & \phantom{1}-1.1 \\
    \bottomrule    
    \end{tabular}
    \label{tab:ihr_cgalore}
    \begin{tablenotes}
    \footnotesize
    \item[a] Difference in Average WER is significant.
    \end{tablenotes}
    \end{threeparttable}
\end{table}

CGaLore is related to  IHR~\cite{vanderEeckt2026inversehessian}, but uses curvature at a different stage. IHR applies a post-hoc correction to the final parameter update $\Delta \bm W$ after adaptation, i.e., the inverse Hessian correction from Eq. \eqref{eq:kfac_proj} is applied to $\Delta \bm W$ learned on the new task which is then added to $\bm W$; whereas CGaLore uses old-task curvature during training to guide $\bm P$ and $\bm R$. Table \ref{tab:ihr_cgalore} shows the performance of CGaLore and GaLore on the first and second adaptation of Exp. 2 when combined with IHR. When applying IHR, we use a hyper-parameter $\lambda \in [0,1]$ that controls how far the final update is rotated toward the inverse-Hessian-corrected direction: $\lambda=0$ leaves the original update unchanged, $\lambda=1$ applies the full IHR correction. We tune $\lambda$ on the validation sets after the first adaptation. The update to which Eq. \eqref{eq:kfac_proj} is applied is normalized to equal $\Delta \bm W$'s original norm. 

Table~\ref{tab:ihr_cgalore} shows that IHR is a strong post-hoc correction for GaLore, substantially reducing the gap to CGaLore. However, IHR does not make CGaLore redundant: in Exp.~2 after the first adaptation, GaLore+IHR reaches 16.58 WER, whereas CGaLore obtains 16.33 and CGaLore+IHR further improves to 16.09. After the second adaptation, the gap between GaLore and CGaLore when both are combined with IHR becomes even larger, with CGaLore + IHR obtaining an average WER of 16.86, further increasing the gap with baselines from Table \ref{tab:results}. This suggests that curvature-guided subspace selection during training can provide a better learning--retention trade-off than correcting the final update only after training. At the same time, the best results often come from combining CGaLore with a mild IHR correction, indicating that CGaLore and IHR are complementary rather than mutually exclusive.

\subsection{Left, Right, and Two-Sided Projection}
\label{sec:ablation}

\begin{table}
    \centering
    \caption{Effect of projection structure in CGaLore on the first adaptation of Exp.~1. ENG is used for KFAC estimation.}
    \begin{threeparttable}
    \begin{tabular}{lcc}
    \toprule
    \multirow{2}{*}{\textbf{Variant}} & \multicolumn{2}{c}{\textbf{Average}} \\
    \cmidrule(lr){2-3}
     & \textbf{WER}$\downarrow$ & \textbf{BWT}$\uparrow$ \\
    \midrule
    Full two-sided, $r=215$ & 10.00\tnote{a} & -0.3 \\
    Left only, $r=38$      & 11.40\tnote{a} & -2.2 \\
    Right only, $r=38$     & \phantom{1}{9.92} & {-0.2} \\
    \bottomrule
    \end{tabular}
    \label{tab:projection_ablation}
    \begin{tablenotes}
    \footnotesize
    \item[a] Difference in Average WER is significant.
    \end{tablenotes}
    \end{threeparttable}
\end{table}

Table~\ref{tab:projection_ablation} compares the two-sided projection with one-sided variants on the first adaptation of Exp.~1, using only ENG for KFAC. To keep the optimizer-state memory approximately matched, we use $r=215$ for the two-sided projection and $r=38$ for the one-sided projections. The results suggest that constraining the input-side subspace is most important in this setting: right projection performs similarly to the full two-sided projection, whereas left projection degrades BWT. This indicates that, for the evaluated accent-adaptation tasks, preventing interference in input/activation-side directions is more important than constraining the output side alone.

\section{Discussion}
CGaLore requires old-task data only to estimate KFAC factors, and Sec.~\ref{sec:kfac_data} shows that this can be done with only a few minutes of speech per task. Moreover, the resulting factors also reduce forgetting on held-out old tasks not used for KFAC estimation. After KFAC has been computed, the old data can be discarded; CGaLore stores curvature factors rather than replay examples. This is a key difference with rehearsal-based methods,  which store a subset of the past data to replay during adaptation to new tasks: they must keep this data for replay during adaptation and typically require sufficiently diverse stored data as  overfitting on the memory remains a risk~\cite{rehearsal_review}. This is particularly limiting for foundation models, where $\T_0$ may contain many tasks and highly diverse data.

CGaLore also differs from recent GaLore extensions~\cite{wang-etal-2025-continual,cheng2026continuous}, which rely on information accumulated during the original training process, which is usually unavailable when adapting a released ASR foundation model. In contrast, CGaLore requires only a small amount of old-task data for KFAC estimation to adapt the model. Compared with PECL methods such as CSSVD~\cite{cssvd} or BiLoRA~\cite{bilora}, CGaLore reduces forgetting through the optimizer trajectory rather than by restricting adaptation to a fixed parameter-efficient update form.

CGaLore leaves several directions for future work. First, unlike recent GaLore extensions~\cite{zhang2024qgalore, lotus}, CGaLore does not automatically determine when each layer should update its projection bases $\bm P$ and $\bm R$. A layer-wise update mechanism that accounts for both current-task gradient drift and old-task KFAC curvature could further reduce the cost of SVD while preserving performance. Second, CGaLore currently uses the same rank $r$ for each layer. A curvature-aware rank allocation strategy could assign larger subspaces to layers that can be adapted safely and smaller subspaces to less relevant or more interference-prone layers, improving the learning--retention trade-off under the same memory budget. Finally, our experiments are limited to a single 367M-parameter ASR model and two-task adaptation sequences. Future work should evaluate CGaLore on larger models and longer task sequences, as well as extend it to foundation-model adaptation in other domains such as NLP or vision.

\section{Conclusion}
We propose CGaLore, an extension of GaLore for memory-efficient continual adaptation of ASR foundation models. CGaLore uses old-task Kronecker-factored approximate curvature (KFAC) estimates to select the GaLore projection bases after first preconditioning the current-task gradient with old-task KFAC estimates, biasing the optimizer subspace toward directions that remain useful for the new task while reducing interference with previous tasks. Experiments on two CL scenarios show that CGaLore substantially reduces forgetting compared with GaLore and improves the learning--retention trade-off over parameter-efficient CL baselines, while preserving GaLore's practical GPU memory efficiency. Our analyses further show that CGaLore can work with only a few minutes of speech per task for KFAC estimation, can also reduce forgetting on held-out old tasks not used to compute KFAC, is not highly sensitive to the projection update interval, and can be combined with post-hoc inverse-Hessian correction for additional gains. These results show that old-task curvature can guide low-rank optimizer subspaces during training, enabling speech foundation models to adapt efficiently while preserving previously learned capabilities. 

\section{Acknowledgments}
Research supported by Research Foundation Flanders (FWO) under grant S004923N of the SBO programme.

\bibliographystyle{ieeetr}
\bibliography{main}

\end{document}